\documentclass{naw-latex}
\usepackage[english]{babel}
\usepackage{graphicx}
\usepackage{xcolor,colortbl}
\usepackage{tikz}
\usetikzlibrary{chains}
\usepackage{algorithm}
\usepackage[noend]{algorithmic}

\newcommand{\LINEIF}[2]{\STATE\algorithmicif\ {#1}\ \algorithmicthen\ {#2}}

\makeatletter
\renewcommand{\ALC@linenosize}{\tiny}
\makeatother
\usepackage{listings}
\usepackage[pdftitle={Binary Pebbling Algorithms for In-Place Reversal of One-Way Hash Chains},pdfauthor={Berry Schoenmakers},bookmarks=false]{hyperref}
\hypersetup{colorlinks=true, urlcolor=blue, linkcolor=blue, citecolor=blue, linktoc = section}

\DeclareMathOperator{\len}{len}
\newcommand{\Z}{\mathbb{Z}}
\newcommand{\tf}[1]{\mbox{$\tfrac{#1}{2}$}}
\newcommand{\assign}{\leftarrow}
\newcommand{\eat}[1]{\mbox{\rm pop}_{#1}}
\makeatletter
\def\La{L\kern -.36em{\sbox \z@ T\vbox to\ht \z@ {\hbox {\check@mathfonts \fontsize \sf@size \z@ \math@fontsfalse \selectfont A}\vss }}}
\makeatother
\definecolor{LightCyan}{rgb}{0.88,1,1}
\newcommand{\red}  [1]{\textcolor[cmyk]{0.00, 1.00, 1.00, 0.25}{#1}}
\newcommand{\blue} [1]{\textcolor[cmyk]{1.00, 1.00, 0.00, 0.25}{#1}}
\newcommand{\green}[1]{\textcolor[cmyk]{1.00, 0.00, 1.00, 0.25}{#1}}

\bibitem{ACP+17}{J.~Alwen, B.~Chen, K.~Pietrzak, L.~Reyzin, and S.~Tessaro.
Scrypt is maximally memory-hard.
In {\em Advances in Cryptology---{EUROCRYPT '}17}, {\em LNCS} 10212:33--62, Springer, 2017.}

\bibitem{BBC+94}{J.-P. Boly, A.~Bosselaers, R.~Cramer, R.~Michelsen, S.~Mj{\o}lsnes, F.~Muller,
  T.~Pedersen, B.~Pfitzmann, P.~de Rooij, B.~Schoenmakers, M.~Schunter,
  L.~Vall{\'{e}}e, and M.~Waidner.
The ESPRIT Project CAFE---High Security Digital Payment Systems.
In {\em Computer Security---ESORICS 94}, {\em LNCS} 875:217--230. Springer, 1994.}

\bibitem{CJ02}{D.~Coppersmith and M.~Jakobsson.
Almost optimal hash sequence traversal.
In {\em Financial Cryptography 2002}, {\em LNCS} 2357:102--119. Springer, 2002.}

\bibitem{FO89}{P.~Flajolet and A.~Odlyzko.
Random mapping statistics.
In {\em Advances in Cryptology---{EUROCRYPT '}89}, {\em LNCS} 434:329--354. Springer, 1989.}

\bibitem{Gri92}{A.~Griewank.
Achieving logarithmic growth of temporal and spatial complexity in reverse automatic differentiation.
In {\em Optimization Methods and Software}, 1(1):35--54, 1992.}

\bibitem{GW08}{A.~Griewank and A.~Walther.
{\em Evaluating Derivatives: Principles and Techniques of Algorithmic Differentiation}. Second edition, SIAM, 2008.}

\bibitem{GPRS96}{J.~Grimm, L.~Potter, and N.~Rostaing-Schmidt.
Optimal time and minimum space-time product for reversing a certain class of programs.
In {\em Computational Differentiation--Techniques, Applications, and Tools}, pp.\ 95--106. SIAM, 1996.}

\bibitem{HN00}{J.~H{\aa}stad and M.~N{\"a}slund.
Key feedback mode: a keystream generator with provable security.
In {\em First Modes of Operation Workshop}, Baltimore, Maryland, USA, October 2000. NIST.}

\bibitem{Hur05}{G.~Hurlbert.
Recent progress in graph pebbling.
{\em Graph Theory Notes of New York}, 49:25--37, 2005.}

\bibitem{IR01}{G.~Itkis and L.~Reyzin.
Forward-secure signatures with optimal signing and verifying.
In {\em Advances in Cryptology---{CRYPTO '}01}, {\em LNCS} 2139:332--354. Springer, 2001.}

\bibitem{Jak02}{M.~Jakobsson.
Fractal hash sequence representation and traversal.
In {\em Proc.\ IEEE International Symposium on Information Theory (ISIT '02)}, p.\ 437. IEEE, 2002.
Full version \href{https://eprint.iacr.org/2002/001/}{\tt eprint.iacr.org/2002/001}.}

\bibitem{Lam81}{L.~Lamport.
Password authentication with insecure communication.
{\em Communications of the ACM}, 24(11):770--772, 1981.}

\bibitem{Lev85}{L.~Levin.
One-way function and pseudorandom generators.
In {\em Proc.\ 17th Symposium on Theory of Computing (STOC '85)},
  pp.\ 363--365. ACM, 1985.}

\bibitem{Mer87}{R.~Merkle.
A digital signature based on a conventional encryption function.
In {\em Advances in Cryptology---{CRYPTO '}87}, {\em LNCS} 293:369--378. Springer, 1987.}

\bibitem{Nak08}{S.~Nakamoto.
Bitcoin: A peer-to-peer electronic cash system. October 31, 2008.
See \href{https://bitcoin.org/bitcoin.pdf}{\tt bitcoin.org/bitcoin.pdf}.}

\bibitem{Nor13}{J.~Nordstr{\"{o}}m.
Pebble games, proof complexity, and time-space trade-offs.
{\em Logical Methods in Computer Science}, 9(3:15):1--63, 2013.}

\bibitem{Ped96}{T.~P. Pedersen.
Electronic payments of small amounts.
In {\em Security Protocols}, {\em LNCS} 1189:59--68. Springer, 1996.}

\bibitem{Per13}{K.~Perumalla.
{\em Introduction to Reversible Computing}.
CRC Press, 2013.}

\bibitem{RS96}{R.~L. Rivest and A.~Shamir.
Payword and micromint: Two simple micropayment schemes.
In {\em Security Protocols}, {\em LNCS} 1189:69--87, Springer, 1996.}

\bibitem{Sch16}{B.~Schoenmakers.
Explicit optimal binary pebbling for one-way hash chain reversal.
In {\em Financial Cryptography 2016}, {\em LNCS} 9603:299–-320. Springer, 2016.
Sample code in Python, Java, C at \href{https://www.win.tue.nl/~berry/pebbling/}{\tt www.win.tue.nl/\~{}berry/pebbling}.}

\bibitem{YSEL09}{D.~H. Yum, J.~W. Seo, S. Eom, and P.~J. Lee.
Single-layer fractal hash chain traversal with almost optimal complexity.
In {\em Topics in Cryptology---{CT-RSA} '09}, {\em LNCS} 5473:325--339. Springer, 2009.}

\begin{document}

\pagestyle{plain}

\StartArtikel[Titel={\LARGE Binary Pebbling Algorithms for In-Place Reversal of One-Way Hash Chains},
%
          AuteurA={Berry Schoenmakers},
          AdresA={Dept of Mathematics \& Computer Science\crlf TU Eindhoven},
          EmailA={\href{mailto:berry@win.tue.nl}{berry@win.tue.nl}},
          AuteurC={\hfill Nieuw Archief voor Wiskunde},
          AdresC={\hfill 5th series, vol.\ 18, no.\ 3, pp.\ 199--204\crlf \hspace*{3cm} \hfill September 2017},
          EmailC={\hfill \href{http://www.nieuwarchief.nl/}{www.nieuwarchief.nl}},
          kolommen={2}
	  ]

\pagestyle{plain}

\StartLeadIn
The resourcefulness of people working in the area of cryptology is striking. A wonderfully diverse set of professionals capable of generating an almost endless stream of new, intriguing research problems and, fortunately---driven by the ever faster changing playing field of our information society and infrastructure under attack---capable of finding many ingenious and compelling solutions as well.

Unsurprisingly, mathematics plays a central role in cryptology along with computer science. The algorithmic problem chosen as the topic of this paper has a unique motivation in cryptography. At the core of its solution lies an intricate fractal structure which turns out to have a very nice and simple characterization.
\StopLeadIn

The problem is formulated in terms of a length-preserving one-way function $f$. A concrete example is the classical Davies-Meyer one-way function constructed from a block cipher such as AES:
\[ f:\left\{\begin{array}{rcl}
\{0,1\}^{128} & \longrightarrow & \{0,1\}^{128} \\
x & \longmapsto & \textrm{AES}_{x}(\mathbf{0}).
\end{array}\right.\]
That is, $f(x)$ is computed as an AES encryption of the trivial all-zero message $\mathbf{0}$ under the key $x$, which can obviously be done efficiently. On the other hand, recovering $x$ from $f(x)$ is tantamount to recovering an AES key given a single plaintext-ciphertext pair, which is assumed to be computationally hard. Therefore, $f$ is called a {\bf one-way function}, as it is easy to evaluate but hard to invert.
Block ciphers like AES are normally used for symmetric encryption to provide confidentiality, whereas one-way functions like $f$ are often used for asymmetric authentication, e.g., in the construction of digital signature schemes.

\onderwerp*{One-way chains}
Back in 1981, Lamport (the ``\La'' in \LaTeX) proposed an elegant asymmetric identification scheme which operates in terms of one-way chains \cite{Lam81}. A {\bf one-way chain} is the sequence formed by the successive iterates of $f$ for a given value.
For example, in a client-server setting, the client may apply $f$ four times for a randomly chosen 128-bit seed value $x_0$ to obtain a length-4 chain:
\[ x_0 \stackrel{f}{\longrightarrow} x_1 \stackrel{f}{\longrightarrow} x_2 \stackrel{f}{\longrightarrow} x_3 \stackrel{f} {\longrightarrow} x_4.\]
Lamport's {\bf identification scheme} then operates as follows. At the start, the client registers itself securely with the server, as a result of which the server associates the endpoint $x_4$ with the client. Depending on the details, this registration step may be rather involved. However, from now on the client may identify itself securely to the server simply by releasing the next preimage on the chain. In the first round of identification, the client releases preimage $x_3$, and the server checks this value by testing if $f(x_3)=x_4$ holds. An eavesdropper obtaining $x_3$ cannot impersonate the client later on because the next round the server will demand a preimage for $x_3$. At this stage, preimage $x_2$ satisfying $f(x_2)=x_3$ is known only to the client; it is not even known to the server yet, which is why the scheme provides {\em asymmetric} authentication.

One-way chains and variations thereof are often referred to as {\bf hash chains} since cryptographic hash functions such as SHA-256 are commonly used as alternatives for $f$.
Hash chains are fundamental to many constructions in cryptography, and even to some forms of cryptanalysis (e.g., rainbow tables).
Bitcoin's blockchain  \cite{Nak08} is probably the best-known example of a hash chain nowadays---but note that blockchains are costly to generate due to the additional ``proof of work'' requirement for the hash values linking successive blocks. Hash chains are also used in digital signature schemes required to be quantum secure, building on work by Merkle from 1979 \cite{Mer87}. Incidentally, Merkle attributes the use of iterated functions to Winternitz. However, Winternitz's idea is to use only one preimage on a length-$n$ chain, basically to securely encode an integer in the set $\{0,\ldots,n-1\}$, whereas Lamport's idea is to use all of the $n$ preimages. The CAFE phone-tick scheme \cite[Section~3.5]{BBC+94} (see also \cite{Ped96}) and later micropayment schemes (e.g., PayWord \cite{RS96}) actually combine these two ideas. In the case of phone-ticks, the caller releases the endpoint of a chain at the start of a call; at each tick, the caller simply releases the next preimage (as in Lamport's scheme) to pay for continuing the call. After the call ends, the phone company only needs to keep the last preimage released by the caller to claim the amount due (as in Winternitz's encoding).

\onderwerp*{Security of one-way chains}
The use of a cryptographic hash function to create a one-way chain is overkill, however. A function like SHA-256 is not just one-way but is also designed to compress bit strings of practically unlimited length, and related to this, SHA-256 is required to be collision-resistant as well. For the security of a one-way chain, $f$ should be one-way, that is, given $y$
in the range of $f$ it must be hard to find any $x$ such that $f(x)=y$. Or rather, as recognized in \cite{Lev85,Ped96}, $f$ should necessarily be {\bf one-way on its iterates}, which says that, for a length-$n$ chain, given an $n$th iterate image $y$ (in the range of $f^n$) it must be hard to find any $x$ such that $f(x)=y$.

Viewing $f$ as a random function (as in the random oracle model for hash functions), it follows that finding such a preimage $x$ takes $2^{128}/n$ time approximately. If $n=1$ this is simply the problem of inverting $f$, which can only be solved by making random guesses for $x$; on each attempt one succeeds with probability $1/2^{128}$. For $n>1$, however, one should not guess randomly. First, observe that the set of $n$th iterate images $y$ (range of $f^n$) is much smaller than $\{0,1\}^{128}$. In fact,
the expected number of $n$th iterate images $y$ is equal to $(1-\tau_n)2^{128}$, where $\tau_0=0$, $\tau_n = e^{-1+\tau_{n-1}}$ for $n\geq1$ \cite[Theorem 2(v)]{FO89}.
To take advantage of the given that $y$ is not just any image but an $n$th iterate image, start with a random guess $x_0$ and then check if $x_1=f(x_0)$ happens to match $y$. Next, compute $x_2=f(x_1)$ and again test for equality with $y$, and continue to do so until $x_n$ is reached. The overall probability of hitting $y$ and thus obtaining a preimage of $y$ as well works out as $n/2^{128}$ approximately. Hence, even for very long chains of length $n=2^{32}$, say, the security level is still $2^{96}$. See also \cite[Theorem~3]{HN00} for a further analysis.

\onderwerp*{Pebbling algorithms}
The above provides a solid basis for Jakobsson's wonderful idea of using efficient pebbling algorithms to make Lamport's scheme practical even for very long chains \cite{Jak02}. Naive implementations would render Lamport's scheme completely impractical: both (i) computing $x_{n-1}=f^{n-1}(x_0)$ to perform the first round of identification, then computing $x_{n-2}=f^{n-2}(x_0)$ from scratch, and so on, and (ii) storing all of $x_0, x_1, \ldots, x_{n-2}, x_{n-1}$ to perform each round of identification instantly, are out of the question. The crux of Jakobsson's pebbling algorithm is to achieve a good {\bf space-time trade-off}: for chains of length $n=2^k$, Jakobsson's algorithm stores $O(\log n)$ hash values throughout, and the maximum number of hashes performed in any round of identification is $O(\log n)$ as well.

Each hash value stored is associated with a {\bf pebble}. For a length-16 chain, 5 pebbles are initially arranged as follows, which is typical of a
{\bf binary pebbling algorithm}:
\[\begin{tikzpicture}[start chain, node distance=-0.4mm]
\node [on chain] {$\stackrel{x_0}{\bullet}$};
\node [on chain] {$\stackrel{x_1}{\cdot}$};
\node [on chain] {$\stackrel{x_2}{\cdot}$};
\node [on chain] {$\stackrel{x_3}{\cdot}$};
\node [on chain] {$\stackrel{x_4}{\cdot}$};
\node [on chain] {$\stackrel{x_5}{\cdot}$};
\node [on chain] {$\stackrel{x_6}{\cdot}$};
\node [on chain] {$\stackrel{x_7}{\cdot}$};
\node [on chain] {$\stackrel{x_8}{\bullet}$};
\node [on chain] {$\stackrel{x_9}{\cdot}$};
\node [on chain] {$\stackrel{x_{10}}{\cdot}$};
\node [on chain] {$\stackrel{x_{11}}{\cdot}$};
\node [on chain] {$\stackrel{x_{12}}{\bullet}$};
\node [on chain] {$\stackrel{x_{13}}{\cdot}$};
\node [on chain] {$\stackrel{x_{14}}{\bullet}$};
\node [on chain] {$\stackrel{x_{15}}{\bullet}$};
\end{tikzpicture}\]
The general pattern is that starting from the rightmost pebble, the distance to the next pebble doubles each time.
From this initial arrangement, the first two elements $x_{15}$ and $x_{14}$ of the reverse of $\{x_0,x_1,\ldots,x_{15}\}$ can be output directly.
For the third element $x_{13}$ we need to apply $f$ once to recompute it from $x_{12}$. The fourth element $x_{12}$ can be output again
without any effort.

To produce $x_{11}$, something interesting happens. Because $f$ is one-way, the only sensible option is to recompute it from $x_8$ as $x_{11}=f^3(x_8)$. But while doing so, the value of $x_{10}=f^2(x_8)$ is also stored for later use. Hence, just before $x_{11}$ is output, the pebbles are arranged as follows:
\[\begin{tikzpicture}[start chain, node distance=-0.4mm]
\node [on chain] {$\stackrel{x_0}{\bullet}$};
\node [on chain] {$\stackrel{x_1}{\cdot}$};
\node [on chain] {$\stackrel{x_2}{\cdot}$};
\node [on chain] {$\stackrel{x_3}{\cdot}$};
\node [on chain] {$\stackrel{x_4}{\cdot}$};
\node [on chain] {$\stackrel{x_5}{\cdot}$};
\node [on chain] {$\stackrel{x_6}{\cdot}$};
\node [on chain] {$\stackrel{x_7}{\cdot}$};
\node [on chain] {$\stackrel{x_8}{\bullet}$};
\node [on chain] {$\stackrel{x_9}{\cdot}$};
\node [on chain] {$\stackrel{x_{10}}{\bullet}$};
\node [on chain] {$\stackrel{x_{11}}{\bullet}$};
\phantom{\node [on chain] {$\stackrel{x_{12}}{\bullet}$};
\node [on chain] {$\stackrel{x_{13}}{\cdot}$};
\node [on chain] {$\stackrel{x_{14}}{\bullet}$};
\node [on chain] {$\stackrel{x_{15}}{\bullet}$};}
\end{tikzpicture}\]
Proceeding this way and computing outputs just-in-time, the {\bf rushing} binary pebbling algorithm $R_k$ is obtained:
\[ \begin{array}{ll}
R_0(x) = \textrm{output } x \\
R_k(x) = R_{k-1}(f^{2^{k-1}}(x)); R_{k-1}(x).
\end{array}\]
The reader may check that $R_k(x)$ outputs the sequence
\[ f^*_k(x) = \{ f^i(x) \}_{i=0}^{2^k-1} \]
in reverse, using $k 2^{k-1}$ hashes in total. In addition, the storage requirements are low: $R_k(x)$ needs to store $x$ for the recursive call $R_{k-1}(x)$ later on, which leads to a maximum of $k+1$ values stored (pebbles) at any moment.

The only drawback is that $R_k$ in the worst case requires time exponential in $k$ between producing successive outputs. Removing these slow rounds is exactly what makes the problem non-trivial. That is, we seek a way to reverse $f^*_k(x)$ satisfying the performance constraints of using $O(k)$ storage (pebbles) and using $O(k)$ applications of $f$ (hashes) between producing {\em any} two successive outputs. To study this problem we introduce a specific framework for binary pebbling algorithms that operate in rounds.

At this point we like to mention that there are many similar notions of ``pebbling'' in the literature. In particular, pebbling games (see, e.g., \cite{Nor13}) are somewhat related, and have recently been used in the context of cryptography to prove memory-hardness of certain hash functions \cite{ACP+17}. Graph pebbling is another well-known problem (see, e.g., \cite{Hur05}). Reversible computing (see, e.g., \cite{Per13}) gives rise to even more uses of pebbling (a.k.a.\ ``checkpointing'', see below). As discussed in \cite{Sch16}, however,
the specific worst-case constraint limiting the number of hashes per round is unique to the cryptographic setting, starting with the work in \cite{IR01,Jak02}.

\begin{figure*}
\begin{center}
\begin{tikzpicture}[thick, scale=0.19]
\draw[thick] (0,31) -- (0,63);
\draw (0.65,63) node {$x$};
\draw[thick,>>->>] (0,57) to[bend left, in=180] (12,48) to[bend right, out=0, in=145] (31,31);
\draw[dotted] (0,31) -- (31,31);
\draw[thick] (0,31) -- (0,0) -- (15,15);
\draw[thick,>>->>] (0,26) to[bend left, in=180] (6,22) to[bend right, out=0] (15,15);
\draw[dotted] (0,15) -- (5.5,15); \draw (7.5,15) node {$P_{k{-}1}$}; \draw[dotted] (9.5,15) -- (15,15);
\draw[thick] (16,31) -- (16,16) -- (23,23);
\draw[thick,>>->>] (16,28) to[bend left, in=180] (20,25) to[bend right, out=0] (23,23);
\draw[dotted] (16,23) -- (17.5,23); \draw (19.25,23) node {$P_{k{-}2}$}; \draw[dotted] (21.25,23) -- (23,23);
\draw[thick] (24,31) -- (24,24) -- (27,27);
\draw[thick,>->] (24,30) to[bend left, in=180] (26,28) to[bend right, out=0] (27,27);
\draw[dotted] (24,27) -- (27,27);
\draw[thick] (28,31) -- (28,28) -- (29,29);
\draw[thick,>->] (28,30) to[bend left, in=180] (29,29);
\draw[dotted] (28,29) -- (29,29);
\draw[thick] (30,31) -- (30,30);
\draw[dotted] (0,0) -- (31,31);
\draw (25,55) node[align=center] {\begin{tabular}{l}
{\bf Initial stage}: \\
- set $y_i=f^{2^k-2^i}(x)$, \\
\phantom{- }for $i=k,\ldots,0$, \\
\phantom{- }using $t_{k,r}$ hashes \\
\phantom{- }in round $r \in [1,2^k)$.
\end{tabular}};
\draw (26,6) node[align=center] {\begin{tabular}{l}
{\bf Output stage}: \\
- output $y_0$ in round $r=2^k$; \\
- run $P_{i-1}(y_i)$ in parallel, \\
\phantom{- }for $i=1,\ldots,k$, \\
\phantom{- }in rounds $r \in (2^k,2^{k+1})$.
\end{tabular}};
\draw (0,31) node {$\bullet$};
\draw (1.2,31.9) node {$y_k$};
\draw (16,45) -- (16,31);
\draw (16,31) node {$\bullet$};
\draw (18,31.9) node {$y_{k{-}1}$};
\draw (24,40.3) -- (24,31);
\draw (24,31) node {$\bullet$};
\draw (26,31.9) node {$y_{k{-}2}$};
\draw (28,37) -- (28,31);
\draw (28,31) node {$\bullet$};
\draw (30,33.9) -- (30,31);
\draw (30,31) node {$\bullet$};
\draw (32.2,31.9) node {$y_0$};
\draw (31,31) node {$\bullet$};
\draw (-0.5,63) node {$-$};
\draw (-0.5,31) node {$-$};
\draw (-0.5,0) node {$-$};
\draw (-6.5,63) node {$r=1$};
\draw (-6.5,31) node {$r=2^k$};
\draw (-6.5,0) node {$r=2^{k{+}1}{-}1$};
\end{tikzpicture}
\hspace*{3mm}
\begin{tikzpicture}[scale=0.185, baseline=-12]
\draw[ultra thin] (0,1) -- (0,61);
\draw[ultra thin] (16,17) -- (16,32);
\draw[ultra thin] (24,25) -- (24,31.5);
\draw[ultra thin] (28,29) -- (28,31.25);
\draw[ultra thin] (20,21) -- (20,23.67);
\draw[ultra thin] (12,13) -- (12,15.29);
\draw[ultra thin] (8,9) -- (8,15.86);
\draw[ultra thin] (4,5) -- (4,7.67);
\draw (-1.2,62.7) node {{\scriptsize $r$}};
\draw (-6,63) node {\scriptsize $T_4$};
\draw (-4,63) node {\scriptsize $S_4$};
\draw (22,56) node[align=center] {rushing \\ pebbler $P_4$};
\foreach \x in {61,59,...,1} {\draw (-1.2,\x) node {{\tiny \pgfmathparse{(63-\x)/2} \pgfmathprintnumber{\pgfmathresult}}};}
\draw (-4,61) node {\scriptsize 1};
\draw (-6,61) node {\scriptsize 0};
\draw (0,61) node {$\bullet$};
\draw (-4,59) node {\scriptsize 1};
\draw (-6,59) node {\scriptsize 0};
\draw (0,59) node {$\bullet$};
\draw (-4,57) node {\scriptsize 1};
\draw (-6,57) node {\scriptsize 0};
\draw (0,57) node {$\bullet$};
\draw (-4,55) node {\scriptsize 1};
\draw (-6,55) node {\scriptsize 0};
\draw (0,55) node {$\bullet$};
\draw (-4,53) node {\scriptsize 1};
\draw (-6,53) node {\scriptsize 0};
\draw (0,53) node {$\bullet$};
\draw (-4,51) node {\scriptsize 1};
\draw (-6,51) node {\scriptsize 0};
\draw (0,51) node {$\bullet$};
\draw (-4,49) node {\scriptsize 1};
\draw (-6,49) node {\scriptsize 0};
\draw (0,49) node {$\bullet$};
\draw (-4,47) node {\scriptsize 1};
\draw (-6,47) node {\scriptsize 0};
\draw (0,47) node {$\bullet$};
\draw (-4,45) node {\scriptsize 1};
\draw (-6,45) node {\scriptsize 0};
\draw (0,45) node {$\bullet$};
\draw (-4,43) node {\scriptsize 1};
\draw (-6,43) node {\scriptsize 0};
\draw (0,43) node {$\bullet$};
\draw (-4,41) node {\scriptsize 1};
\draw (-6,41) node {\scriptsize 0};
\draw (0,41) node {$\bullet$};
\draw (-4,39) node {\scriptsize 1};
\draw (-6,39) node {\scriptsize 0};
\draw (0,39) node {$\bullet$};
\draw (-4,37) node {\scriptsize 1};
\draw (-6,37) node {\scriptsize 0};
\draw (0,37) node {$\bullet$};
\draw (-4,35) node {\scriptsize 1};
\draw (-6,35) node {\scriptsize 0};
\draw (0,35) node {$\bullet$};
\draw (-4,33) node {\scriptsize 1};
\draw (-6,33) node {\scriptsize 15};
\draw (0,33) node {$\bullet$};
\draw[semithick,->] (0,33) to (29.7,31.3);
\draw (-4,31) node {\scriptsize 5};
\draw (-6,31) node {\scriptsize $W_4$};
\draw[dotted] (0,31) -- (30,31);
\draw (0,31) node {$\bullet$};
\draw (16,31) node {$\bullet$};
\draw (24,31) node {$\bullet$};
\draw (28,31) node {$\bullet$};
\draw (30,31) node {$\bullet$};
\draw (-4,29) node {\scriptsize 4};
\draw (-6,29) node {\scriptsize 1};
\draw (0,29) node {$\bullet$};
\draw (16,29) node {$\bullet$};
\draw (24,29) node {$\bullet$};
\draw[semithick,->] (24,29) to (25.7,27.3);
\draw (28,29) node {$\bullet$};
\draw (-4,27) node {\scriptsize 4};
\draw (-6,27) node {\scriptsize 0};
\draw (0,27) node {$\bullet$};
\draw (16,27) node {$\bullet$};
\draw (24,27) node {$\bullet$};
\draw[dotted] (24,27) -- (26,27);
\draw (26,27) node {$\bullet$};
\draw (-4,25) node {\scriptsize 3};
\draw (-6,25) node {\scriptsize 3};
\draw (0,25) node {$\bullet$};
\draw (16,25) node {$\bullet$};
\draw[semithick,->] (16,25) to (21.7,23.3);
\draw (24,25) node {$\bullet$};
\draw (-4,23) node {\scriptsize 4};
\draw (-6,23) node {\scriptsize 0};
\draw (0,23) node {$\bullet$};
\draw (16,23) node {$\bullet$};
\draw[dotted] (16,23) -- (22,23);
\draw (20,23) node {$\bullet$};
\draw (22,23) node {$\bullet$};
\draw (-4,21) node {\scriptsize 3};
\draw (-6,21) node {\scriptsize 1};
\draw (0,21) node {$\bullet$};
\draw (16,21) node {$\bullet$};
\draw[semithick,->] (16,21) to (17.7,19.3);
\draw (20,21) node {$\bullet$};
\draw (-4,19) node {\scriptsize 3};
\draw (-6,19) node {\scriptsize 0};
\draw (0,19) node {$\bullet$};
\draw (16,19) node {$\bullet$};
\draw[dotted] (16,19) -- (18,19);
\draw (18,19) node {$\bullet$};
\draw (-4,17) node {\scriptsize 2};
\draw (-6,17) node {\scriptsize 7};
\draw (0,17) node {$\bullet$};
\draw[semithick,->] (0,17) to (13.7,15.3);
\draw (16,17) node {$\bullet$};
\draw (-4,15) node {\scriptsize 4};
\draw (-6,15) node {\scriptsize 0};
\draw (0,15) node {$\bullet$};
\draw[dotted] (0,15) -- (14,15);
\draw (8,15) node {$\bullet$};
\draw (12,15) node {$\bullet$};
\draw (14,15) node {$\bullet$};
\draw (-4,13) node {\scriptsize 3};
\draw (-6,13) node {\scriptsize 1};
\draw (0,13) node {$\bullet$};
\draw (8,13) node {$\bullet$};
\draw[semithick,->] (8,13) to (9.7,11.3);
\draw (12,13) node {$\bullet$};
\draw (-4,11) node {\scriptsize 3};
\draw (-6,11) node {\scriptsize 0};
\draw (0,11) node {$\bullet$};
\draw (8,11) node {$\bullet$};
\draw[dotted] (8,11) -- (10,11);
\draw (10,11) node {$\bullet$};
\draw (-4,9) node {\scriptsize 2};
\draw (-6,9) node {\scriptsize 3};
\draw (0,9) node {$\bullet$};
\draw[semithick,->] (0,9) to (5.7,7.3);
\draw (8,9) node {$\bullet$};
\draw (-4,7) node {\scriptsize 3};
\draw (-6,7) node {\scriptsize 0};
\draw (0,7) node {$\bullet$};
\draw[dotted] (0,7) -- (6,7);
\draw (4,7) node {$\bullet$};
\draw (6,7) node {$\bullet$};
\draw (-4,5) node {\scriptsize 2};
\draw (-6,5) node {\scriptsize 1};
\draw (0,5) node {$\bullet$};
\draw[semithick,->] (0,5) to (1.7,3.3);
\draw (4,5) node {$\bullet$};
\draw (-4,3) node {\scriptsize 2};
\draw (-6,3) node {\scriptsize 0};
\draw (0,3) node {$\bullet$};
\draw[dotted] (0,3) -- (2,3);
\draw (2,3) node {$\bullet$};
\draw (-4,1) node {\scriptsize 1};
\draw (-6,1) node {\scriptsize 0};
\draw (0,1) node {$\bullet$};
\end{tikzpicture}
\end{center}
\caption{(left) Binary pebbler $P_k(x)$ for schedule $T_k=\{ t_{k,r} \}_{r=1}^{2^k-1}$ satisfying $\sum_{r=1}^{2^k-1} t_{k,r} =2^k-1$. (right) Schedule $T_4$, work $W_4$, and storage $S_4$ for rushing pebbler $P_4$ in rounds $r=1$ to $r=31$.
Bullets represent stored values (pebbles), rightwards arrows represent hashing, vertical lines represent copying.}
\label{figure:binary-pebbler}
\end{figure*}
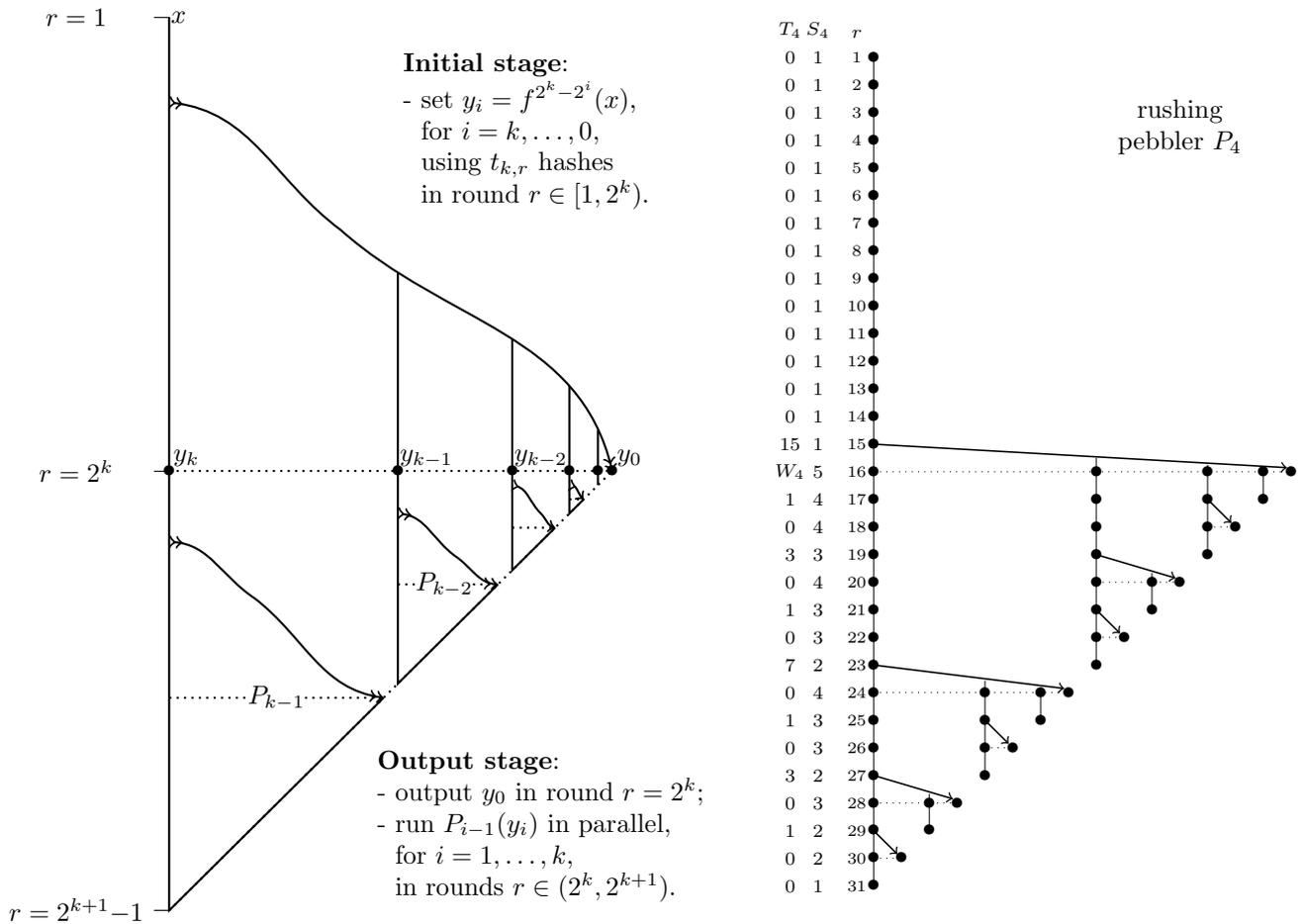

\onderwerp*{Framework for binary pebbling}
For $k\geq0$, pebbler $P_k(x)$ will be defined as an algorithm that runs for $2^{k+1}-1$ rounds
in total, and outputs $f^*_k(x)$ in reverse in its last $2^k$ rounds. It is essential
that we include the initial $2^k-1$ rounds (in which no outputs are produced) as an integral part of
$P_k(x)$, as this allows for a fully recursive definition and analysis of binary pebbling.
In fact, in terms of a given {\bf schedule} $T_k=\{ t_{k,r} \}_{r=1}^{2^k-1}$, which fixes the number of hashes for each initial round,
a {\bf binary pebbler} $P_k(x)$ is completely specified by the recursive definition given in
Figure~\ref{figure:binary-pebbler}. This means, for example, that $P_0(x)$ runs for one round only outputting $y_0=x$ itself,
and that $P_1(x)$ will run for three rounds, performing $t_{1,1}=1$ hash in its first round, outputting $y_0=f(x)$ in its
second round, and outputting $y_1=x$ in its last round. In general, $P_k(x)$ computes $f^{2^k-1}(x)$ using exactly $2^k-1$ hashes in total in its
initial stage, storing only the values $y_k,\ldots,y_0$ along the way.
Running pebblers $P_{k-1},\ldots,P_0$ in parallel in the output stage means that pebblers take turns
to execute for one round each, where the order in which this happens within a round is irrelevant. It is not hard to prove that
in every round exactly one of the pebblers running in parallel will be in its first output round, and that the
sequence of outputs is always equal to $f^*_k(x)$.

Schedule $T_k$ specifies the number of hashes for the initial stage of $P_k$.
To analyze the {\bf work} done by $P_k$ in its output stage, we let sequence $W_k$ of length $2^k-1$ denote the
number of hashes performed by $P_k$ in each of its last $2^k-1$ rounds---noting that by definition
no hashes are performed by $P_k$ in round $2^k$. The following recurrence relation for $W_k$ will be useful throughout:
\[W_0 = \{\},\quad  W_k = T_{k-1}+W_{k-1} \;\; \| \;\; \{ 0 \} \;\; \| \;\; W_{k-1},\]
where $T_{k-1}+W_{k-1}$ denotes elementwise addition of $T_{k-1}$ and $W_{k-1}$ and $\|$ concatenation of sequences ($+$ takes precedence over $\|$).

To analyze the {\bf storage} needed by $P_k$ the number of hash values stored by $P_k$ will be counted for each round.
We let sequence $S_k = \{s_{k,r}\}_{r=1}^{2^{k+1}-1}$ denote the total storage used by $P_k$ at the start of each round.
For instance, $s_{k,1}=1$ as $P_k$ only stores $x$ at the start, and $s_{k,2^k}=k+1$
as $P_k$ stores $y_0,\ldots,y_k$ at the start of round $2^k$ independent of schedule $T_k$.

The rushing pebbler $P_k$ corresponding to $R_k$ introduced above is obtained by taking schedule $T_k$ with $t_{k,2^k-1} = 2^k-1$ and $t_{k,r}=0$ elsewhere.
Rushing pebbler $P_4$ is illustrated in Figure~\ref{figure:binary-pebbler} in our framework for binary pebbling. The storage $S_4$ is minimal throughout, but for the work $W_4$ there are big peaks: e.g., in round~23, in total 7 hashes are performed, while the pebbler is idle in all even rounds.

\begin{figure*}[t]
\begin{tikzpicture}[scale=0.1433]
\draw[ultra thin] (0,1) -- (0,61);
\draw[ultra thin] (16,17) -- (16,45);
\draw[ultra thin] (24,25) -- (24,37);
\draw[ultra thin] (28,29) -- (28,33);
\draw[ultra thin] (20,21) -- (20,25);
\draw[ultra thin] (12,13) -- (12,17);
\draw[ultra thin] (8,9) -- (8,21);
\draw[ultra thin] (4,5) -- (4,9);
\draw (-1.2,62.7) node {{\scriptsize $r$}};
\draw (-6,63) node {\scriptsize $T_4$};
\draw (-4,63) node {\scriptsize $S_4$};
\draw (22,56) node[align=center] {speed-1 \\ pebbler $P_4$};
\foreach \x in {61,59,...,1} {\draw (-1.35,\x) node {{\tiny \pgfmathparse{(63-\x)/2} \pgfmathprintnumber{\pgfmathresult}}};}
\draw (-4,61) node {\scriptsize 1};
\draw (-6,61) node {\scriptsize 1};
\draw (0,61) node {$\bullet$};
\draw[semithick,->] (0,61) to (1.7,59.3);
\draw (-4,59) node {\scriptsize 2};
\draw (-6,59) node {\scriptsize 1};
\draw (0,59) node {$\bullet$};
\draw (2,59) node {$\bullet$};
\draw[semithick,->] (2,59) to (3.7,57.3);
\draw (-4,57) node {\scriptsize 2};
\draw (-6,57) node {\scriptsize 1};
\draw (0,57) node {$\bullet$};
\draw (4,57) node {$\bullet$};
\draw[semithick,->] (4,57) to (5.7,55.3);
\draw (-4,55) node {\scriptsize 2};
\draw (-6,55) node {\scriptsize 1};
\draw (0,55) node {$\bullet$};
\draw (6,55) node {$\bullet$};
\draw[semithick,->] (6,55) to (7.7,53.3);
\draw (-4,53) node {\scriptsize 2};
\draw (-6,53) node {\scriptsize 1};
\draw (0,53) node {$\bullet$};
\draw (8,53) node {$\bullet$};
\draw[semithick,->] (8,53) to (9.7,51.3);
\draw (-4,51) node {\scriptsize 2};
\draw (-6,51) node {\scriptsize 1};
\draw (0,51) node {$\bullet$};
\draw (10,51) node {$\bullet$};
\draw[semithick,->] (10,51) to (11.7,49.3);
\draw (-4,49) node {\scriptsize 2};
\draw (-6,49) node {\scriptsize 1};
\draw (0,49) node {$\bullet$};
\draw (12,49) node {$\bullet$};
\draw[semithick,->] (12,49) to (13.7,47.3);
\draw (-4,47) node {\scriptsize 2};
\draw (-6,47) node {\scriptsize 1};
\draw (0,47) node {$\bullet$};
\draw (14,47) node {$\bullet$};
\draw[semithick,->] (14,47) to (15.7,45.3);
\draw (-4,45) node {\scriptsize 2};
\draw (-6,45) node {\scriptsize 1};
\draw (0,45) node {$\bullet$};
\draw (16,45) node {$\bullet$};
\draw[semithick,->] (16,45) to (17.7,43.3);
\draw (-4,43) node {\scriptsize 3};
\draw (-6,43) node {\scriptsize 1};
\draw (0,43) node {$\bullet$};
\draw (16,43) node {$\bullet$};
\draw (18,43) node {$\bullet$};
\draw[semithick,->] (18,43) to (19.7,41.3);
\draw (-4,41) node {\scriptsize 3};
\draw (-6,41) node {\scriptsize 1};
\draw (0,41) node {$\bullet$};
\draw (16,41) node {$\bullet$};
\draw (20,41) node {$\bullet$};
\draw[semithick,->] (20,41) to (21.7,39.3);
\draw (-4,39) node {\scriptsize 3};
\draw (-6,39) node {\scriptsize 1};
\draw (0,39) node {$\bullet$};
\draw (16,39) node {$\bullet$};
\draw (22,39) node {$\bullet$};
\draw[semithick,->] (22,39) to (23.7,37.3);
\draw (-4,37) node {\scriptsize 3};
\draw (-6,37) node {\scriptsize 1};
\draw (0,37) node {$\bullet$};
\draw (16,37) node {$\bullet$};
\draw (24,37) node {$\bullet$};
\draw[semithick,->] (24,37) to (25.7,35.3);
\draw (-4,35) node {\scriptsize 4};
\draw (-6,35) node {\scriptsize 1};
\draw (0,35) node {$\bullet$};
\draw (16,35) node {$\bullet$};
\draw (24,35) node {$\bullet$};
\draw (26,35) node {$\bullet$};
\draw[semithick,->] (26,35) to (27.7,33.3);
\draw (-4,33) node {\scriptsize 4};
\draw (-6,33) node {\scriptsize 1};
\draw (0,33) node {$\bullet$};
\draw (16,33) node {$\bullet$};
\draw (24,33) node {$\bullet$};
\draw (28,33) node {$\bullet$};
\draw[semithick,->] (28,33) to (29.7,31.3);
\draw (-4,31) node {\scriptsize 5};
\draw (-6,31) node {\scriptsize $W_4$};
\draw[dotted] (0,31) -- (30,31);
\draw (0,31) node {$\bullet$};
\draw (16,31) node {$\bullet$};
\draw (24,31) node {$\bullet$};
\draw (28,31) node {$\bullet$};
\draw (30,31) node {$\bullet$};
\draw (-4,29) node {\scriptsize 4};
\draw (-6,29) node {\scriptsize 3};
\draw (0,29) node {$\bullet$};
\draw[semithick,->] (0,29) to (1.7,27.3);
\draw (16,29) node {$\bullet$};
\draw[semithick,->] (16,29) to (17.7,27.3);
\draw (24,29) node {$\bullet$};
\draw[semithick,->] (24,29) to (25.7,27.3);
\draw (28,29) node {$\bullet$};
\draw (-4,27) node {\scriptsize 6};
\draw (-6,27) node {\scriptsize 2};
\draw (0,27) node {$\bullet$};
\draw (2,27) node {$\bullet$};
\draw[semithick,->] (2,27) to (3.7,25.3);
\draw (16,27) node {$\bullet$};
\draw (18,27) node {$\bullet$};
\draw[semithick,->] (18,27) to (19.7,25.3);
\draw (24,27) node {$\bullet$};
\draw[dotted] (24,27) -- (26,27);
\draw (26,27) node {$\bullet$};
\draw (-4,25) node {\scriptsize 5};
\draw (-6,25) node {\scriptsize 2};
\draw (0,25) node {$\bullet$};
\draw (4,25) node {$\bullet$};
\draw[semithick,->] (4,25) to (5.7,23.3);
\draw (16,25) node {$\bullet$};
\draw (20,25) node {$\bullet$};
\draw[semithick,->] (20,25) to (21.7,23.3);
\draw (24,25) node {$\bullet$};
\draw (-4,23) node {\scriptsize 5};
\draw (-6,23) node {\scriptsize 1};
\draw (0,23) node {$\bullet$};
\draw (6,23) node {$\bullet$};
\draw[semithick,->] (6,23) to (7.7,21.3);
\draw (16,23) node {$\bullet$};
\draw[dotted] (16,23) -- (22,23);
\draw (20,23) node {$\bullet$};
\draw (22,23) node {$\bullet$};
\draw (-4,21) node {\scriptsize 4};
\draw (-6,21) node {\scriptsize 2};
\draw (0,21) node {$\bullet$};
\draw (8,21) node {$\bullet$};
\draw[semithick,->] (8,21) to (9.7,19.3);
\draw (16,21) node {$\bullet$};
\draw[semithick,->] (16,21) to (17.7,19.3);
\draw (20,21) node {$\bullet$};
\draw (-4,19) node {\scriptsize 5};
\draw (-6,19) node {\scriptsize 1};
\draw (0,19) node {$\bullet$};
\draw (8,19) node {$\bullet$};
\draw (10,19) node {$\bullet$};
\draw[semithick,->] (10,19) to (11.7,17.3);
\draw (16,19) node {$\bullet$};
\draw[dotted] (16,19) -- (18,19);
\draw (18,19) node {$\bullet$};
\draw (-4,17) node {\scriptsize 4};
\draw (-6,17) node {\scriptsize 1};
\draw (0,17) node {$\bullet$};
\draw (8,17) node {$\bullet$};
\draw (12,17) node {$\bullet$};
\draw[semithick,->] (12,17) to (13.7,15.3);
\draw (16,17) node {$\bullet$};
\draw (-4,15) node {\scriptsize 4};
\draw (-6,15) node {\scriptsize 0};
\draw (0,15) node {$\bullet$};
\draw[dotted] (0,15) -- (14,15);
\draw (8,15) node {$\bullet$};
\draw (12,15) node {$\bullet$};
\draw (14,15) node {$\bullet$};
\draw (-4,13) node {\scriptsize 3};
\draw (-6,13) node {\scriptsize 2};
\draw (0,13) node {$\bullet$};
\draw[semithick,->] (0,13) to (1.7,11.3);
\draw (8,13) node {$\bullet$};
\draw[semithick,->] (8,13) to (9.7,11.3);
\draw (12,13) node {$\bullet$};
\draw (-4,11) node {\scriptsize 4};
\draw (-6,11) node {\scriptsize 1};
\draw (0,11) node {$\bullet$};
\draw (2,11) node {$\bullet$};
\draw[semithick,->] (2,11) to (3.7,9.3);
\draw (8,11) node {$\bullet$};
\draw[dotted] (8,11) -- (10,11);
\draw (10,11) node {$\bullet$};
\draw (-4,9) node {\scriptsize 3};
\draw (-6,9) node {\scriptsize 1};
\draw (0,9) node {$\bullet$};
\draw (4,9) node {$\bullet$};
\draw[semithick,->] (4,9) to (5.7,7.3);
\draw (8,9) node {$\bullet$};
\draw (-4,7) node {\scriptsize 3};
\draw (-6,7) node {\scriptsize 0};
\draw (0,7) node {$\bullet$};
\draw[dotted] (0,7) -- (6,7);
\draw (4,7) node {$\bullet$};
\draw (6,7) node {$\bullet$};
\draw (-4,5) node {\scriptsize 2};
\draw (-6,5) node {\scriptsize 1};
\draw (0,5) node {$\bullet$};
\draw[semithick,->] (0,5) to (1.7,3.3);
\draw (4,5) node {$\bullet$};
\draw (-4,3) node {\scriptsize 2};
\draw (-6,3) node {\scriptsize 0};
\draw (0,3) node {$\bullet$};
\draw[dotted] (0,3) -- (2,3);
\draw (2,3) node {$\bullet$};
\draw (-4,1) node {\scriptsize 1};
\draw (-6,1) node {\scriptsize 0};
\draw (0,1) node {$\bullet$};
\end{tikzpicture}
\hspace*{1mm}
\begin{tikzpicture}[scale=0.1433]
\draw[ultra thin] (0,1) -- (0,61);
\draw[ultra thin] (16,17) -- (16,39);
\draw[ultra thin] (24,25) -- (24,35);
\draw[ultra thin] (28,29) -- (28,33);
\draw[ultra thin] (20,21) -- (20,25);
\draw[ultra thin] (12,13) -- (12,17);
\draw[ultra thin] (8,9) -- (8,19);
\draw[ultra thin] (4,5) -- (4,9);
\draw (-1.2,62.7) node {{\scriptsize $r$}};
\draw (-6,63) node {\scriptsize $T_4$};
\draw (-4,63) node {\scriptsize $S_4$};
\draw (22,56) node[align=center] {speed-2 \\ pebbler $P_4$};
\foreach \x in {61,59,...,1} {\draw (-1.35,\x) node {{\tiny \pgfmathparse{(63-\x)/2} \pgfmathprintnumber{\pgfmathresult}}};}
\draw (-4,61) node {\scriptsize 1};
\draw (-6,61) node {\scriptsize 0};
\draw (0,61) node {$\bullet$};
\draw (-4,59) node {\scriptsize 1};
\draw (-6,59) node {\scriptsize 0};
\draw (0,59) node {$\bullet$};
\draw (-4,57) node {\scriptsize 1};
\draw (-6,57) node {\scriptsize 0};
\draw (0,57) node {$\bullet$};
\draw (-4,55) node {\scriptsize 1};
\draw (-6,55) node {\scriptsize 0};
\draw (0,55) node {$\bullet$};
\draw (-4,53) node {\scriptsize 1};
\draw (-6,53) node {\scriptsize 0};
\draw (0,53) node {$\bullet$};
\draw (-4,51) node {\scriptsize 1};
\draw (-6,51) node {\scriptsize 0};
\draw (0,51) node {$\bullet$};
\draw (-4,49) node {\scriptsize 1};
\draw (-6,49) node {\scriptsize 0};
\draw (0,49) node {$\bullet$};
\draw (-4,47) node {\scriptsize 1};
\draw (-6,47) node {\scriptsize 2};
\draw (0,47) node {$\bullet$};
\draw[semithick,->] (0,47) to (3.7,45.3);
\draw (-4,45) node {\scriptsize 2};
\draw (-6,45) node {\scriptsize 2};
\draw (0,45) node {$\bullet$};
\draw (4,45) node {$\bullet$};
\draw[semithick,->] (4,45) to (7.7,43.3);
\draw (-4,43) node {\scriptsize 2};
\draw (-6,43) node {\scriptsize 2};
\draw (0,43) node {$\bullet$};
\draw (8,43) node {$\bullet$};
\draw[semithick,->] (8,43) to (11.7,41.3);
\draw (-4,41) node {\scriptsize 2};
\draw (-6,41) node {\scriptsize 2};
\draw (0,41) node {$\bullet$};
\draw (12,41) node {$\bullet$};
\draw[semithick,->] (12,41) to (15.7,39.3);
\draw (-4,39) node {\scriptsize 2};
\draw (-6,39) node {\scriptsize 2};
\draw (0,39) node {$\bullet$};
\draw (16,39) node {$\bullet$};
\draw[semithick,->] (16,39) to (19.7,37.3);
\draw (-4,37) node {\scriptsize 3};
\draw (-6,37) node {\scriptsize 2};
\draw (0,37) node {$\bullet$};
\draw (16,37) node {$\bullet$};
\draw (20,37) node {$\bullet$};
\draw[semithick,->] (20,37) to (23.7,35.3);
\draw (-4,35) node {\scriptsize 3};
\draw (-6,35) node {\scriptsize 2};
\draw (0,35) node {$\bullet$};
\draw (16,35) node {$\bullet$};
\draw (24,35) node {$\bullet$};
\draw[semithick,->] (24,35) to (27.7,33.3);
\draw (-4,33) node {\scriptsize 4};
\draw (-6,33) node {\scriptsize 1};
\draw (0,33) node {$\bullet$};
\draw (16,33) node {$\bullet$};
\draw (24,33) node {$\bullet$};
\draw (28,33) node {$\bullet$};
\draw[semithick,->] (28,33) to (29.7,31.3);
\draw (-4,31) node {\scriptsize 5};
\draw (-6,31) node {\scriptsize $W_4$};
\draw[dotted] (0,31) -- (30,31);
\draw (0,31) node {$\bullet$};
\draw (16,31) node {$\bullet$};
\draw (24,31) node {$\bullet$};
\draw (28,31) node {$\bullet$};
\draw (30,31) node {$\bullet$};
\draw (-4,29) node {\scriptsize 4};
\draw (-6,29) node {\scriptsize 1};
\draw (0,29) node {$\bullet$};
\draw (16,29) node {$\bullet$};
\draw (24,29) node {$\bullet$};
\draw[semithick,->] (24,29) to (25.7,27.3);
\draw (28,29) node {$\bullet$};
\draw (-4,27) node {\scriptsize 4};
\draw (-6,27) node {\scriptsize 2};
\draw (0,27) node {$\bullet$};
\draw (16,27) node {$\bullet$};
\draw[semithick,->] (16,27) to (19.7,25.3);
\draw (24,27) node {$\bullet$};
\draw[dotted] (24,27) -- (26,27);
\draw (26,27) node {$\bullet$};
\draw (-4,25) node {\scriptsize 4};
\draw (-6,25) node {\scriptsize 1};
\draw (0,25) node {$\bullet$};
\draw (16,25) node {$\bullet$};
\draw (20,25) node {$\bullet$};
\draw[semithick,->] (20,25) to (21.7,23.3);
\draw (24,25) node {$\bullet$};
\draw (-4,23) node {\scriptsize 4};
\draw (-6,23) node {\scriptsize 2};
\draw (0,23) node {$\bullet$};
\draw[semithick,->] (0,23) to (3.7,21.3);
\draw (16,23) node {$\bullet$};
\draw[dotted] (16,23) -- (22,23);
\draw (20,23) node {$\bullet$};
\draw (22,23) node {$\bullet$};
\draw (-4,21) node {\scriptsize 4};
\draw (-6,21) node {\scriptsize 3};
\draw (0,21) node {$\bullet$};
\draw (4,21) node {$\bullet$};
\draw[semithick,->] (4,21) to (7.7,19.3);
\draw (16,21) node {$\bullet$};
\draw[semithick,->] (16,21) to (17.7,19.3);
\draw (20,21) node {$\bullet$};
\draw (-4,19) node {\scriptsize 4};
\draw (-6,19) node {\scriptsize 2};
\draw (0,19) node {$\bullet$};
\draw (8,19) node {$\bullet$};
\draw[semithick,->] (8,19) to (11.7,17.3);
\draw (16,19) node {$\bullet$};
\draw[dotted] (16,19) -- (18,19);
\draw (18,19) node {$\bullet$};
\draw (-4,17) node {\scriptsize 4};
\draw (-6,17) node {\scriptsize 1};
\draw (0,17) node {$\bullet$};
\draw (8,17) node {$\bullet$};
\draw (12,17) node {$\bullet$};
\draw[semithick,->] (12,17) to (13.7,15.3);
\draw (16,17) node {$\bullet$};
\draw (-4,15) node {\scriptsize 4};
\draw (-6,15) node {\scriptsize 0};
\draw (0,15) node {$\bullet$};
\draw[dotted] (0,15) -- (14,15);
\draw (8,15) node {$\bullet$};
\draw (12,15) node {$\bullet$};
\draw (14,15) node {$\bullet$};
\draw (-4,13) node {\scriptsize 3};
\draw (-6,13) node {\scriptsize 1};
\draw (0,13) node {$\bullet$};
\draw (8,13) node {$\bullet$};
\draw[semithick,->] (8,13) to (9.7,11.3);
\draw (12,13) node {$\bullet$};
\draw (-4,11) node {\scriptsize 3};
\draw (-6,11) node {\scriptsize 2};
\draw (0,11) node {$\bullet$};
\draw[semithick,->] (0,11) to (3.7,9.3);
\draw (8,11) node {$\bullet$};
\draw[dotted] (8,11) -- (10,11);
\draw (10,11) node {$\bullet$};
\draw (-4,9) node {\scriptsize 3};
\draw (-6,9) node {\scriptsize 1};
\draw (0,9) node {$\bullet$};
\draw (4,9) node {$\bullet$};
\draw[semithick,->] (4,9) to (5.7,7.3);
\draw (8,9) node {$\bullet$};
\draw (-4,7) node {\scriptsize 3};
\draw (-6,7) node {\scriptsize 0};
\draw (0,7) node {$\bullet$};
\draw[dotted] (0,7) -- (6,7);
\draw (4,7) node {$\bullet$};
\draw (6,7) node {$\bullet$};
\draw (-4,5) node {\scriptsize 2};
\draw (-6,5) node {\scriptsize 1};
\draw (0,5) node {$\bullet$};
\draw[semithick,->] (0,5) to (1.7,3.3);
\draw (4,5) node {$\bullet$};
\draw (-4,3) node {\scriptsize 2};
\draw (-6,3) node {\scriptsize 0};
\draw (0,3) node {$\bullet$};
\draw[dotted] (0,3) -- (2,3);
\draw (2,3) node {$\bullet$};
\draw (-4,1) node {\scriptsize 1};
\draw (-6,1) node {\scriptsize 0};
\draw (0,1) node {$\bullet$};
\end{tikzpicture}
\hspace*{1mm}
\begin{tikzpicture}[scale=0.1433]
\draw[ultra thin] (0,1) -- (0,61);
\draw[ultra thin] (16,17) -- (16,37);
\draw[ultra thin] (24,25) -- (24,33);
\draw[ultra thin] (28,29) -- (28,31.67);
\draw[ultra thin] (20,21) -- (20,24);
\draw[ultra thin] (12,13) -- (12,16);
\draw[ultra thin] (8,9) -- (8,18);
\draw[ultra thin] (4,5) -- (4,8);
\draw (-1.2,62.7) node {{\scriptsize $r$}};
\draw (-6,63) node {\scriptsize $T_4$};
\draw (-4,63) node {\scriptsize $S_4$};
\draw (22,56) node[align=center] {optimal \\ pebbler $P_4$};
\foreach \x in {61,59,...,1} {\draw (-1.35,\x) node {{\tiny \pgfmathparse{(63-\x)/2} \pgfmathprintnumber{\pgfmathresult}}};}
\draw (-4,61) node {\scriptsize 1};
\draw (-6,61) node {\scriptsize 0};
\draw (0,61) node {$\bullet$};
\draw (-4,59) node {\scriptsize 1};
\draw (-6,59) node {\scriptsize 0};
\draw (0,59) node {$\bullet$};
\draw (-4,57) node {\scriptsize 1};
\draw (-6,57) node {\scriptsize 0};
\draw (0,57) node {$\bullet$};
\draw (-4,55) node {\scriptsize 1};
\draw (-6,55) node {\scriptsize 0};
\draw (0,55) node {$\bullet$};
\draw (-4,53) node {\scriptsize 1};
\draw (-6,53) node {\scriptsize 0};
\draw (0,53) node {$\bullet$};
\draw (-4,51) node {\scriptsize 1};
\draw (-6,51) node {\scriptsize 0};
\draw (0,51) node {$\bullet$};
\draw (-4,49) node {\scriptsize 1};
\draw (-6,49) node {\scriptsize 0};
\draw (0,49) node {$\bullet$};
\draw (-4,47) node {\scriptsize 1};
\draw (-6,47) node {\scriptsize 2};
\draw (0,47) node {$\bullet$};
\draw[semithick,->] (0,47) to (3.7,45.3);
\draw (-4,45) node {\scriptsize 2};
\draw (-6,45) node {\scriptsize 2};
\draw (0,45) node {$\bullet$};
\draw (4,45) node {$\bullet$};
\draw[semithick,->] (4,45) to (7.7,43.3);
\draw (-4,43) node {\scriptsize 2};
\draw (-6,43) node {\scriptsize 1};
\draw (0,43) node {$\bullet$};
\draw (8,43) node {$\bullet$};
\draw[semithick,->] (8,43) to (9.7,41.3);
\draw (-4,41) node {\scriptsize 2};
\draw (-6,41) node {\scriptsize 1};
\draw (0,41) node {$\bullet$};
\draw (10,41) node {$\bullet$};
\draw[semithick,->] (10,41) to (11.7,39.3);
\draw (-4,39) node {\scriptsize 2};
\draw (-6,39) node {\scriptsize 2};
\draw (0,39) node {$\bullet$};
\draw (12,39) node {$\bullet$};
\draw[semithick,->] (12,39) to (15.7,37.3);
\draw (-4,37) node {\scriptsize 2};
\draw (-6,37) node {\scriptsize 2};
\draw (0,37) node {$\bullet$};
\draw (16,37) node {$\bullet$};
\draw[semithick,->] (16,37) to (19.7,35.3);
\draw (-4,35) node {\scriptsize 3};
\draw (-6,35) node {\scriptsize 2};
\draw (0,35) node {$\bullet$};
\draw (16,35) node {$\bullet$};
\draw (20,35) node {$\bullet$};
\draw[semithick,->] (20,35) to (23.7,33.3);
\draw (-4,33) node {\scriptsize 3};
\draw (-6,33) node {\scriptsize 3};
\draw (0,33) node {$\bullet$};
\draw (16,33) node {$\bullet$};
\draw (24,33) node {$\bullet$};
\draw[semithick,->] (24,33) to (29.7,31.3);
\draw (-4,31) node {\scriptsize 5};
\draw (-6,31) node {\scriptsize $W_4$};
\draw[dotted] (0,31) -- (30,31);
\draw (0,31) node {$\bullet$};
\draw (16,31) node {$\bullet$};
\draw (24,31) node {$\bullet$};
\draw (28,31) node {$\bullet$};
\draw (30,31) node {$\bullet$};
\draw (-4,29) node {\scriptsize 4};
\draw (-6,29) node {\scriptsize 1};
\draw (0,29) node {$\bullet$};
\draw (16,29) node {$\bullet$};
\draw (24,29) node {$\bullet$};
\draw[semithick,->] (24,29) to (25.7,27.3);
\draw (28,29) node {$\bullet$};
\draw (-4,27) node {\scriptsize 4};
\draw (-6,27) node {\scriptsize 1};
\draw (0,27) node {$\bullet$};
\draw (16,27) node {$\bullet$};
\draw[semithick,->] (16,27) to (17.7,25.3);
\draw (24,27) node {$\bullet$};
\draw[dotted] (24,27) -- (26,27);
\draw (26,27) node {$\bullet$};
\draw (-4,25) node {\scriptsize 4};
\draw (-6,25) node {\scriptsize 2};
\draw (0,25) node {$\bullet$};
\draw (16,25) node {$\bullet$};
\draw (18,25) node {$\bullet$};
\draw[semithick,->] (18,25) to (21.7,23.3);
\draw (24,25) node {$\bullet$};
\draw (-4,23) node {\scriptsize 4};
\draw (-6,23) node {\scriptsize 2};
\draw (0,23) node {$\bullet$};
\draw[semithick,->] (0,23) to (3.7,21.3);
\draw (16,23) node {$\bullet$};
\draw[dotted] (16,23) -- (22,23);
\draw (20,23) node {$\bullet$};
\draw (22,23) node {$\bullet$};
\draw (-4,21) node {\scriptsize 4};
\draw (-6,21) node {\scriptsize 2};
\draw (0,21) node {$\bullet$};
\draw (4,21) node {$\bullet$};
\draw[semithick,->] (4,21) to (5.7,19.3);
\draw (16,21) node {$\bullet$};
\draw[semithick,->] (16,21) to (17.7,19.3);
\draw (20,21) node {$\bullet$};
\draw (-4,19) node {\scriptsize 4};
\draw (-6,19) node {\scriptsize 2};
\draw (0,19) node {$\bullet$};
\draw (6,19) node {$\bullet$};
\draw[semithick,->] (6,19) to (9.7,17.3);
\draw (16,19) node {$\bullet$};
\draw[dotted] (16,19) -- (18,19);
\draw (18,19) node {$\bullet$};
\draw (-4,17) node {\scriptsize 4};
\draw (-6,17) node {\scriptsize 2};
\draw (0,17) node {$\bullet$};
\draw (8,17) node {$\bullet$};
\draw (10,17) node {$\bullet$};
\draw[semithick,->] (10,17) to (13.7,15.3);
\draw (16,17) node {$\bullet$};
\draw (-4,15) node {\scriptsize 4};
\draw (-6,15) node {\scriptsize 0};
\draw (0,15) node {$\bullet$};
\draw[dotted] (0,15) -- (14,15);
\draw (8,15) node {$\bullet$};
\draw (12,15) node {$\bullet$};
\draw (14,15) node {$\bullet$};
\draw (-4,13) node {\scriptsize 3};
\draw (-6,13) node {\scriptsize 1};
\draw (0,13) node {$\bullet$};
\draw (8,13) node {$\bullet$};
\draw[semithick,->] (8,13) to (9.7,11.3);
\draw (12,13) node {$\bullet$};
\draw (-4,11) node {\scriptsize 3};
\draw (-6,11) node {\scriptsize 1};
\draw (0,11) node {$\bullet$};
\draw[semithick,->] (0,11) to (1.7,9.3);
\draw (8,11) node {$\bullet$};
\draw[dotted] (8,11) -- (10,11);
\draw (10,11) node {$\bullet$};
\draw (-4,9) node {\scriptsize 3};
\draw (-6,9) node {\scriptsize 2};
\draw (0,9) node {$\bullet$};
\draw (2,9) node {$\bullet$};
\draw[semithick,->] (2,9) to (5.7,7.3);
\draw (8,9) node {$\bullet$};
\draw (-4,7) node {\scriptsize 3};
\draw (-6,7) node {\scriptsize 0};
\draw (0,7) node {$\bullet$};
\draw[dotted] (0,7) -- (6,7);
\draw (4,7) node {$\bullet$};
\draw (6,7) node {$\bullet$};
\draw (-4,5) node {\scriptsize 2};
\draw (-6,5) node {\scriptsize 1};
\draw (0,5) node {$\bullet$};
\draw[semithick,->] (0,5) to (1.7,3.3);
\draw (4,5) node {$\bullet$};
\draw (-4,3) node {\scriptsize 2};
\draw (-6,3) node {\scriptsize 0};
\draw (0,3) node {$\bullet$};
\draw[dotted] (0,3) -- (2,3);
\draw (2,3) node {$\bullet$};
\draw (-4,1) node {\scriptsize 1};
\draw (-6,1) node {\scriptsize 0};
\draw (0,1) node {$\bullet$};
\end{tikzpicture}
\caption{Schedule $T_4$, work $W_4$, storage $S_4$ for three types of binary pebblers $P_4$ in rounds 1--31. \protect\phantom{aaa aaa bbb bbb ccc ccc} \vspace*{-\baselineskip}}
\label{figure:binary-pebbles-P4}
\end{figure*}
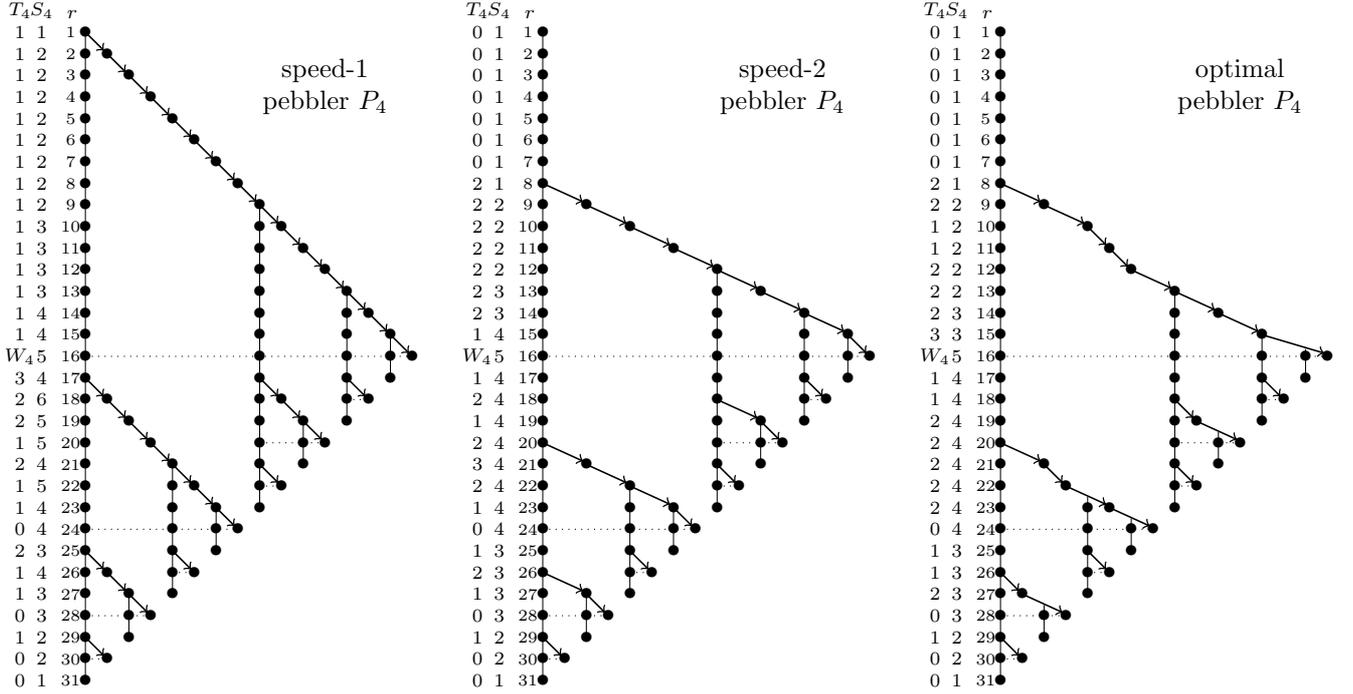

\onderwerp*{Towards optimal solution}
As it turns out, our framework admits a simple solution obtained by taking schedule $T_k = \{ 1 \}_{r=1}^{2^k-1}$, resulting in the {\bf speed-1} pebbler illustrated in Figure~\ref{figure:binary-pebbles-P4}. The above recurrence relation for $W_k$ yields $\max(W_k)=k-1$ for $k\geq1$, and it can also be shown that $\max(S_k)=\max(k+1,2k-2)=O(k)$. The speed-1 pebbler thus achieves the desired asymptotic bounds. For practical purposes, however, further savings are needed to limit the costs as much as possible. E.g., to enable a lightweight client device to identify itself every half hour for a period of three years using a length-$2^{16}$ chain.

Jakobsson's pebbling algorithm \cite{Jak02} provides a clever way to cut storage $\max(S_k)$ in half essentially. Translated to our framework for binary pebbling, the corresponding schedule $T_k$ is obtained by setting $t_{k,r}=0$ for $1\leq r<2^{k-1}$, $t_{k,r}=2$ for $2^{k-1}\leq r<2^k-1$, and $t_{k,2^k-1} = 1$. The pebbler obtained this way is called the {\bf speed-2} pebbler, illustrated in Figure~\ref{figure:binary-pebbles-P4}.

Compared to a speed-1 pebbler, the crucial idea of a speed-2 pebbler is to remain idle for the first half of the initial stage---preventing that too many pebblers are active at the same time---and then make up for this by hashing at double speed in the remaining time.
It can be proved that $\max(W_k)=k-1$ for $k\geq1$ also holds for a speed-2 pebbler, but compared to a speed-1 pebbler storage is now reduced by 50\%,
achieving $\max(S_k)=k+1$. For binary pebbling algorithms, storage $S_k$ of up to $k+1$ hash values is optimal, since this amount of storage is already needed during the first output round $r=2^k$, for any binary pebbler $P_k$. The interesting question is whether the work $\max(W_k)$ can be reduced any further?

\onderwerp*{Optimal binary pebbling}
An elementary analysis yields $\max(W_k)\geq\lceil k/2\rceil$, $k\geq2$, as lower bound for any binary pebbling algorithm (see~\cite[Theorem~2]{Sch16}). So, the best that can be achieved is to reduce the maximum number of hashes for any output round to $k/2$ roughly. The problem of optimally efficient hash chain reversal was extensively studied by Coppersmith and Jakobsson~\cite{CJ02}. They achieved nearly optimal space-time complexity for a complicated pebbling algorithm using $k + \lceil\log_2 (k+1)\rceil$ pebbles and no more than $\lfloor \tfrac{k}{2}\rfloor$ hashes per round. Hence, an excess storage of approximately $\log_2 k$ hash values compared to optimal binary pebbling.

Fortunately, Yum et al.~\cite{YSEL09} observed that a greedy implementation of Jakobsson's original pebbling algorithm already achieves the optimal space-time trade-off for binary pebbling. Their idea is to greedily use up a budget of $\lceil\tfrac{k}{2}\rceil$ hashes per round subject to the constraint that no more than about $k$ hash values are stored at any time. The only drawback of the greedy approach is that no apparent structure is revealed.
\begin{table*}[b]
\caption{Recursive definition of optimal schedule $T_k= \{0\}^{2^{k-1}-1} \,\|\, U_k \,\|\, V_k$ over $\tf1\Z$ (no rounding).
Explicit formula is in this case given by $T_k= \{0\}^{2^{k-1}-1} \,\|\, \{\tfrac{1}{2}(k+1-\len((2r)\bmod 2^{\len(2^k-r)}))\}_{r=2^{k-1}}^{2^k-1}$.}
\begin{center}
\renewcommand{\arraystretch}{1.25}
$\begin{array}{|c|r@{\hspace{0pt}}l|c|r@{\hspace{0pt}}l|} \hline
\multicolumn{3}{|l|}{U_2=\{\tf3\}, U_k = \blue{U_{k-1} + \tf1} \; \| \; \green{\{1\}^{2^{k-3}}}} &
\multicolumn{3}{l|}{V_2=\{\tf3\}, V_k = \blue{U_{k-1} + \tf1} \; \| \; \red{V_{k-1} + \tf1}} \\\hline
U_3 & \blue{2} & \green{1} &
V_3 & \blue{2} & \red{2}\\
U_4 & \blue{\tf5\tf3} & \green{1 1} &
V_4 &  \blue{\tf5\tf3} & \red{\tf{5}\tf{5}}\\
U_5 & \blue{3 2 \tf3\tf3} &\green{1111} &
V_5 & \blue{3 2 \tf3\tf3} &\red{3233} \\
U_6 & \blue{\tf7\tf522\tf3\tf3\tf3\tf3} &\green{11111111} &
V_6 &\blue{\tf7\tf522\tf3\tf3\tf3\tf3} &\red{\tf7\tf522\tf7\tf5\tf7\tf7} \\
U_7 & \blue{43\tf5\tf52222\tf3\tf3\tf3\tf3\tf3\tf3\tf3\tf3}&\green{1111111111111111}&
V_7 & \blue{43\tf5\tf52222\tf3\tf3\tf3\tf3\tf3\tf3\tf3\tf3}&\red{43\tf5\tf5222243\tf5\tf54344} \\\hline
\multicolumn{1}{c}{} & \multicolumn{1}{c}{\blue{\textrm{avg.\ speed\ 2}}} & \multicolumn{1}{c}{\green{\textrm{speed\ 1}}} &
\multicolumn{1}{c}{} & \multicolumn{1}{c}{\blue{\textrm{avg.\ speed\ 2}}} & \multicolumn{1}{c}{\red{\textrm{avg.\ speed\ 3}}}
\end{array}$
\end{center}
\label{table:optimal-recursive}
\end{table*}

In contrast, we have found an explicit, essentially unique solution for optimal binary pebbling, which leads to a complete understanding of the problem and paves the way for fully optimized in-place implementations. As a closed formula, the optimal schedule $T_k$ is obtained by setting $t_{k,r}=0$ for $1\leq r<2^{k-1}$, and setting $t_{k,r}$ to
\[\left\lfloor\tfrac{1}{2}\left((k+r)\bmod 2+k+1-\len((2r)\bmod 2^{\len(2^k-r)})\right)\right\rfloor\]
for $2^{k-1}\leq r< 2^k$, where $\len(n)$ denotes the bit length of nonnegative integer $n$.
Optimal pebbler $P_4$ is illustrated in Figure~\ref{figure:binary-pebbles-P4}, which uses the following optimal schedules:
\[\begin{array}{rcl}
 T_0 &=& \{\}, \\
 T_1 &=& \{1\}, \\
 T_2 &=& \{0,1,2\}, \\
 T_3 &=& \{0,0,0,2,1,2,2\}, \\
 T_4 &=& \{0,0,0,0,0,0,0,2,2,1,1,2,2,2,3\}.
\end{array}\]
In general, an optimal pebbler $P_k$ will use up to $\max(W_k)=\lceil k/2 \rceil$ hashes in any output round. For the optimal pebbler $P_4$ in Figure~\ref{figure:binary-pebbles-P4}, this works out as $\max(W_4)=2$ hashes, compared to the speed-2 pebbler $P_4$ which needs $3$ hashes in output round $r=21$.

The fractal nature of the optimal schedule $T_k$ is revealed by the recursive characterization in terms of sequences $U_k, V_k$ defined in Table~\ref{table:optimal-recursive}. These
sequences are defined over $\tf1 \Z$---rather than over $\Z$ as will ultimately be required for use in a pebbling algorithm. Without rounding of these half-integers, the
optimal schedule satisfies the following {\bf key equation} in terms of sequences $U_k, V_k$, $k\geq2$:
\[ (U_k \;\; \| \;\; V_k) \;\;  + \;\;  (\{0\} \;\; \| \;\; W_{k-1}) \; = \; \{\tfrac{k+1}{2}\}^{2^{k-1}}. \]
This equation basically says that the optimal schedule does not leave any gaps: in each round exactly the maximum number of hashes are performed to meet the lower bound for binary pebbling.

\onderwerp*{Efficient in-place implementations}
Without strict performance requirements, our framework for binary pebbling allows for relatively straightforward implementations. Figure~\ref{figure:python-pebbling} is showcasing
a conceptually simple implementation based on Python generators. For demonstration purposes, we are using MD5 as a 128-bit length-preserving one-way function---MD5 is readily available in Python, also no practical attacks against the one-wayness of MD5 are known to this day.

By exploiting specific properties of the optimal schedule, we will next show how to implement binary pebblers with minimal overhead. In fact, we present {\bf in-place} hash chain reversal algorithms, where the entire state of these algorithms (apart from the hash values) is represented between rounds by a single $k$-bit counter {\em only}. Below, this is shown for Jakobsson's speed-2 pebblers; refer to \cite{Sch16} for further results.

\begin{figure*}
\parbox{6cm}{
\begin{algorithmic}[1]
\ENSURE $r$:
\STATE output $z[0]$
\STATE $c\assign 2^{k+1}-r$
\STATE $i\assign\eat{0}(c)$
\STATE $z[0,i)\assign z[1,i]$ \label{line:shift-z-tail}
\STATE $i\assign i+1$; $c\assign\lfloor c/2 \rfloor$
\STATE $q\assign i-1$
\WHILE{$c\neq0$} \label{line:in-situ-speed-2-while}
    \STATE $z[q]\assign f(z[i])$ \label{line:in-situ-speed-2-clone}
    \LINEIF{$q\neq0$}{$z[q]\assign f(z[q])$}
    \STATE $i\assign i+\eat{0}(c)+\eat{1}(c)$ \label{line:in-situ-speed-2-advance-i}
    \STATE $q\assign i$
\ENDWHILE \label{line:in-situ-speed-2-end-while}
\end{algorithmic}}
\begin{tabular}{p{8mm}p{8mm}p{8mm}p{8mm}p{8mm}p{8mm}p{8mm}p{8mm}p{12mm}}
$c_8$ & $c_7$ & $c_6$ & $c_5$ & $c_4$ & $c_3$ & $c_2$ & $c_1$ & $c_0$ \\
\cellcolor{blue!22}1 & \cellcolor{blue!22}0 & \cellcolor{LightCyan}1 & \cellcolor{red!18}1 & \cellcolor{red!18}0 & \cellcolor{blue!10}1 &\cellcolor{blue!10}0 & \cellcolor{blue!10}0 & \cellcolor{blue!10}0 \\[1mm]
\multicolumn{2}{>{\columncolor{blue!22}}l}{$P_{8_{/y_8/y_7}}^{\textrm{\tiny hashing}}$}&
\cellcolor{LightCyan}$P_{6_{/y_6}}^{\textrm{\tiny idle}}$ &
\multicolumn{2}{>{\columncolor{red!18}}l}{$P_{5_{/y_5/y_4}}^{\textrm{\tiny hashing}}$} &
\multicolumn{4}{>{\columncolor{blue!10}}l}{$P_{3_{/y_3/y_2/y_1/y_0}}^{\textrm{\tiny hashing}}$} \\[1mm]
\\
\cellcolor{blue!22}1 & \cellcolor{blue!22}0 & \cellcolor{LightCyan}1 & \cellcolor{red!18}1 & \cellcolor{red!18}0 & \cellcolor{red!18}0 & \cellcolor{orange!10}1 & \cellcolor{yellow!10} 1 & \cellcolor{magenta!10} 1 \\[1mm]
\multicolumn{2}{>{\columncolor{blue!22}}l}{$P_{8_{/y_8/y_7}}^{\textrm{\tiny hashing}}$}&
\cellcolor{LightCyan}$P_{6_{/y_6}}^{\textrm{\tiny idle}}$ &
\multicolumn{3}{>{\columncolor{red!18}}l}{$P_{5_{/y_5/y_4/y_3}}^{\textrm{\tiny hashing}}$} &
\cellcolor{orange!10}$\!P_{2_{/y_2}}^{\textrm{\tiny idle}}$ &
\cellcolor{yellow!10}$\!P_{1_{/y_1}}^{\textrm{\tiny idle}}$ &
\cellcolor{magenta!10}$\!\!P_{0_{/y_0}}^{\textrm{\tiny hashing}}$ \\[1mm]
$z[8]$ & $z[7]$ & $z[6]$ & $z[5]$ & $z[4]$ & $z[3]$ & $z[2]$ & $z[1]$ & $z[0]$ \\
\\
\multicolumn{9}{r}{\footnotesize $P_{i_{/\textit{hash values}}}^{\textrm{ state}}$: $P_i$ with state and hash values stored in array $z$} \\[1mm]
\multicolumn{9}{r}{\footnotesize $\eat{0}(c)$ / $\eat{1}(c)$: count and remove trailing 0-bits / 1-bits from $c$}
\end{tabular}
\caption{(left) Pseudocode for in-place speed-2 pebbler $P_k(x)$ at output round $r$, $2^k<r<2^{k+1}$. Initially, array $z[0,k)$ satisfies $z[i{-}1]=f^{2^k{-}2^i}(x)$ for $i=1,\ldots,k$. (right) Transition of $P_9$ from round $r=664$ to $r=665$, hence from $c=360=(101101000)_2$ to $c=359=(101100111)_2$.}
\label{figure:state360}
\end{figure*}
We use the following terminology to describe the state of a pebbler $P_k$ (which applies to both speed-2 pebblers and optimal pebblers).
Pebbler $P_k$ is said to be {\bf idle} if it is in rounds $[1,2^{k-1})$,
{\bf hashing} if it is in rounds $[2^{k-1},2^k]$, and {\bf redundant} if it is in rounds $(2^k,2^{k+1})$. An idle pebbler performs no hashes at
all, while a hashing pebbler will perform at least one hash per round, except for round $2^k$ in which $P_k$ outputs its $y_0$ value. The
work for a redundant pebbler $P_k$ is taken over by its child pebblers $P_0,\ldots,P_{k-1}$ during its last $2^k-1$ output rounds.

The important observation is that for each round $r$ the complete state of a pebbler $P_k$ can be deduced quickly from the binary representation of the counter $c=2^{k+1}-r$, which counts down how many rounds are still left. This is illustrated in Figure~\ref{figure:state360} for a speed-2 pebbler $P_k(x)$. The pseudocode shows how to run the pebbler in-place, that is, in such a way that the storage between rounds is limited to a length-$k$ array $z$ of hash values and counter $r$. The information about the states of all pebblers running in parallel is deduced directly from $c$. This information includes which pebblers are present, whether these pebblers are idle or hashing, which hash values have already been computed by a pebbler, and where these are stored in array $z$, etc.

The example in Figure~\ref{figure:state360} shows the details for a $P_9$ pebbler at round $r=664$. Four child pebblers $P_8, P_6, P_5, P_3$ are running in parallel: $P_8$ is hashing and has entries $z[7,8]$ in use, $P_6$ is idle occupying one entry $z[6]$, $P_5$ is hashing and has entries $z[4,5]$ in use. The $P_3$ pebbler has just reached its first output round occupying four entries $z[0,3]$ and outputs its $y_0$ value stored in $z[0]$. Subsequently, this $P_3$ pebbler becomes redundant and is replaced by its child pebblers $P_2,P_1,P_0$, which will each use one entry of array $z$. Entry $z[3]$ has been freed, but is immediately used again by the $P_5$ pebbler, which just reached the point where it starts working on its $y_3$ value.

The schedule for a speed-2 pebbler is integrated in the pseudocode of Figure~\ref{figure:state360}. For optimal pebbling, however, we need to evaluate the formula for the optimal schedule to find the exact number of hashes to be performed by each pebbler. An intuitive way to interpret this formula is explained by means of the following example, cf.\ Figure~\ref{figure:state360}. Consider optimal pebbler $P_9$ at $c=360$ rounds from the end:
\begin{center}
\begin{tabular}{p{3.0mm}p{3.0mm}p{3.0mm}p{3.0mm}p{3.0mm}p{3.0mm}p{3.0mm}p{3.0mm}p{3.0mm}}
$c_8$ & $c_7$ & $c_6$ & $c_5$ & $c_4$ & $c_3$ & $c_2$ & $c_1$ & $c_0$ \\
\cellcolor{blue!22}1 &\cellcolor{blue!22}0 & \cellcolor{blue!22}1 &
\cellcolor{red!18}1 & \cellcolor{red!18}0 & \cellcolor{red!18}1 & \cellcolor{red!18}0 & \cellcolor{red!18}0 & \cellcolor{red!18}0\\
\multicolumn{3}{l}{$P_8^{\textrm{\tiny hashing}}$} &
\multicolumn{5}{l}{$P_5^{\textrm{\tiny hashing}}$}
\end{tabular}
\end{center}
The formula of Table~\ref{table:optimal-recursive} for the optimal schedule (before rounding) partitions the bits of $c$ into the two colored segments as indicated. The underlying rule is as follows. First, all the hashing pebblers are identified, ignoring the rightmost one: this results in two hashing pebblers $P_8$ and $P_5$ (idle pebbler $P_6$ and the rightmost hashing pebbler $P_3$ are ignored). Then, each of these hashing pebblers $P_i$ gets the segment assigned starting at bit $c_i$ and extending to the right. The number of hashes to be performed by each of these hashing pebblers---as given by the formula of the optimal schedule---exactly matches the length of these segments divided by 2. In case of $P_8$ this works out as $\tf3$ hashes, and for $P_5$ we get $\tf6$ hashes, hence
exactly $\tf9$ hashes are used in total for this round.

In general, this rule implies that no more than $\tf{k}$ hashes are performed in any output round of $P_k$.  Moreover, this simple rule will orchestrate the entire computation, ensuring that all intermediate hash values are computed right on time---not too late to fail producing an output on time, and not too early, before another free entry in array $z[0,k)$ becomes available. The optimized implementations in \cite{Sch16} are based on this rule.

\onderwerp*{Lower bound}
Optimal binary pebbling achieves a space-time product of $0.50 k^2$ for a chain of length $n=2^k$. In an upcoming paper with Niels de Vreede, we will show how
to reduce the space-time product to $0.46 k^2$ by means of Fibonacci pebbling (see also \cite{Sch16}) and how to reduce this even further down to just $0.37 k^2$ by more intricate pebbling algorithms. We note that Coppersmith and Jakobsson \cite{CJ02} gave a lower bound of $0.25 k^2$, but whether this bound can be attained is doubtful: the lower bound is derived without taking into account any limits on the number of hashes per output round.

Incidentally, the lower bound of $0.25k^2$ had been found already in a completely different context \cite{GPRS96}, for a similar problem studied in the area of algorithmic (or, automatic, computational) differentiation \cite{GW08}. The lower bound applies to the space-time complexity of so-called {\bf checkpointing} for the reverse (or, adjoint, backward) mode of algorithmic differentiation. In contrast to our case, however, there it is even possible to attain the lower bound \cite{Gri92}. The critical difference is that in the setting of algorithmic differentiation the goal is basically to minimize the {\em total time} for performing this task (or, equivalently, to minimize the {\em amortized time} per output round). This contrasts sharply with the goal in the cryptographic setting, where we want to minimize the {\em worst case time} per output round while performing this task.

\begin{figure*}
\begin{lstlisting}[language=Python, basicstyle=\footnotesize, keywords={yield,lambda,if,else,import,def,for,in}, ndkeywords=, keywordstyle=\bfseries, basewidth=0.45em, columns=fixed, xleftmargin=1cm, frame=single, linewidth=.95\linewidth]
import hashlib, itertools

f = lambda x: hashlib.md5(x).digest()

tR = lambda k,r: 0 if r < 2**k - 1 else 2**k - 1
t1 = lambda k,r: 1
t2 = lambda k,r: 0 if r < 2**(k-1) else 2 if r < 2**k - 1 else 1
tS = lambda k,r: 0 if r < 2**(k-1) else ((k + r) % 2 + k + 1 - ((2 * r) % (2**(2**k - r).bit_length())).bit_length()) // 2

def P(k,x):
    y = [None] * k + [x]
    i = k; g = 0
    for r in range(1, 2**k):
        for _ in range(t(k,r)):
            z = y[i]
            if g == 0: i -= 1; g = 2**i
            y[i] = f(z)
            g -= 1
        yield
    yield y[0]
    for v in itertools.zip_longest(*(P(i-1, y[i]) for i in range(1, k+1))):
        yield next(filter(None, v))

t = eval(input())
k = int(input())
x = f(b'')
for v in P(k,x):
    if v: print(v.hex())
\end{lstlisting}
\caption{Python program for recursive binary pebblers  without any optimizations, cf.\ definition of $P_k(x)$ in Fig.~\ref{figure:binary-pebbler}.
Inputs: {\tt tR/t1/t2/tS} for rushing/speed-1/speed-2/optimal and nonnegative integer {\tt k}.
{\tt P(k,x)} is a Python generator: each evaluation of a {\tt yield} expression corresponds to a round of $P_k(x)$.}
\label{figure:python-pebbling}
\end{figure*}

\completepublications

\end{document}